\documentclass[preprint]{aastex}
\input epsf

\newcommand{\vi}{\mbox{$V\!-\!I$}}

\newcommand{\feh}{\mbox{$[{\rm Fe/H}]$}}
\begin{document}
\title{New Members of the Cluster Family in Nearby Lenticulars 
 \footnote{Based on data obtained at the W.\ M.\ Keck Observatory, which is 
 operated as a scientific partnership among the California Institute of 
 Technology, the University of California and the National Aeronautics and 
 Space Administration.}
}

\author{Jean P. Brodie and S{\o}ren S. Larsen
  \affil{UC Observatories / Lick Observatory, University of California,
         Santa Cruz, CA 95064, USA}
  \email{brodie@ucolick.org and soeren@ucolick.org}
} 

\begin{abstract}

Using spectra obtained with the Keck I telescope we have demonstrated
conclusively that the faint (23$\le$V$\le24$ mag.) and unusually extended
objects we discovered in HST images of the lenticular galaxies, NGC~1023
and NGC~3384, are star clusters associated with their respective
galaxies. In the case of NGC~1023 we were further able to establish that
these objects are old ($\ge$7--8 Gyr), moderately metal-rich
($\feh=-0.58\pm0.24$) and, having a system rotation curve which is very
similar to that of the host galaxy, are associated with the lenticular
disk.  $\alpha$-element to iron abundance ratios are highly supersolar with
[$\alpha$/Fe] between +0.3 and +0.6. With moderately high metallicities and 
luminosities, and effective radii in the range
7--15 pc (compared to the 2--3 pc sizes typical of normal globular and open
clusters), this population of clusters has no known analog in the Milky Way or
elsewhere in the Local Group.
  
\end{abstract}

\keywords{galaxies: star clusters ---
          galaxies: individual (NGC~1023, NGC~3384)}

\section{Introduction}

  During a recent search for globular clusters in the nearby (9.8 Mpc)
lenticular galaxy NGC~1023 we noticed a number of objects with
larger effective radii than normal globular clusters and fainter magnitudes
\citep[][hereafter Paper I]{lb00}. While normal globular clusters have
typical effective (half-light) radii of about 3 pc and follow a roughly
Gaussian magnitude distribution with a peak at $M_V\approx-7.4$
\citep[e.g.][]{har91}, the objects in NGC~1023 were found to have larger
effective radii of about 7--15 pc and were mostly \emph{fainter} than
$M_V=-7$. In Paper I we also noted that the spatial distribution of these
objects seemed to indicate association with the disk of NGC~1023 and that
they had predominantly red colors, indicating high metallicities and ages
greater than a Gyr.  At first glance, the size distributions of
these objects are reminiscent of the few faint ``Palomar'' type
globular clusters in the outer halo of the Milky Way. However, as we shall
show, the spatial and luminosity
distributions as well as kinematics and metallicities of the 
much more numerous faint extended
objects in NGC~1023 are strikingly different, defining a new parameter space
which must be accommodated in cluster formation models. 

  Because of the faintness of these clusters, it is not surprising that
they have not been noticed before in the numerous HST-based studies of
extragalactic globular clusters. With $M_V$ magnitudes fainter than $-7$,
they will generally have apparent magnitudes fainter than $V=24$ even at
the distance of the Virgo or Fornax clusters, and are thus easily missed or
discarded as possible background galaxies. Furthermore, size information is
necessary to recognize them as
different from normal globular clusters.  In \citet{lar01}, a smaller
number of faint, extended objects similar to those in NGC~1023 were noted
in another nearby \citep[11.5 Mpc,][]{sak97} lenticular, NGC~3384, but
their existence was ruled out in two other nearby early-type galaxies,
NGC~3115 (a lenticular) and NGC~3379 (an elliptical).

  In this paper we show that the faint extended objects in NGC~1023 and
NGC~3384 occupy a new region of the size/luminosity/metallicity parameter
space for star clusters.  We present new Keck / LRIS spectra of these
clusters to verify membership with NGC~1023 and NGC~3384 and disprove the
(albeit unlikely) possibility that the objects are unrelated background
galaxies. We also measure radial velocities for individual objects and
examine the hypothesis that they are associated with the disk, in which
case strong signatures of rotation around the center of the galaxy should
be seen.  We then use the co-added spectra of all objects to put
constraints on their ages and metallicities.  We end with a discussion of
the factors which may have influenced the formation and survival of these
clusters.

\section{Data}
\label{sec:data}

  The cluster candidates were selected from the WFPC2 data of Paper I,
where all aspects of the photometry and measurement of object sizes have 
been discussed.  Figure~\ref{fig:brodie.fig1} illustrates the selection criteria: 
The lower panels are color-magnitude diagrams with symbol size proportional 
to the object sizes, while the upper panels show object size vs.\ \vi\
color for objects brighter than $V=24$. The faint extended objects are
easily identifiable in the color-magnitude diagrams, with $\vi \approx 1.3$
and with FWHM values between 1 and 2 pixels, corresponding to effective
radii of 7 -- 15 pc. The boxes indicate the selection criteria for 
objects to be included in the Keck / LRIS slitmasks. 
 
 Although relative positions of objects measured on a single WFPC2 image would 
easily be accurate enough to provide coordinates for slitmask design, 
significant offsets (1--2$\arcsec$) can be present between individual WFPC2 
pointings. Since the objects in NGC~1023 were selected from two WFPC2
pointings, we therefore re-identified the clusters on an image of 
NGC~1023 taken with LRIS during an earlier observing run and used positions 
measured on the LRIS image to provide the final coordinate lists. For
NGC~3384, only one WFPC2 pointing was used for object selection and the
coordinates measured on the WFPC2 image were used directly for the slitmask
design.

  The observations were carried out on Dec 9 and Dec 10, 2001 in multi-slit
mode with the LRIS spectrograph \citep{oke95} on the Keck I
telescope. Spectra were obtained simultaneously with the two arms on LRIS,
using a dichroic that split at a wavelength of $\sim5600$ \AA . On the blue
side a 400 l/mm grism was used, providing a spectral resolution of about
8 \AA\ in the range $\approx$3500--5500 \AA . On the red side we used a 400
l/mm grating blazed at 7500 \AA, covering the range $\sim5700$--9000 \AA\
with a similar resolution. We had intended to use the red spectra, and the
Ca II triplet at $\sim$8500--8700\AA\ in particular, to determine radial
velocities, but the strong atmospheric OH bands made it impossible to
extract useful spectra for these very faint objects at wavelengths
$\ga8000$ \AA. However, for some objects we could measure radial
velocities using the wavelength region $\sim$5700--7000\AA\ on the red
spectra.  The total integration times for NGC~1023 and NGC~3384 were 570
min (9 hours 30 min) and 300 min (5 hours), respectively, divided into 1
hour integrations.

  Spectral reductions were carried out using the SPECRED package in IRAF\footnote{IRAF is distributed by the National Optical Astronomical Observatories, which are operated by the Association of Universities for Research in Astronomy}.
Wavelength calibrations were performed using spectra of calibration
lamps mounted within LRIS, taken immediately after the last exposure of 
each galaxy. For the blue side, comparison with calibration spectra taken 
during the afternoon showed shifts of about 3--4 \AA, probably due to 
flexure in the instrument.  For the red spectra the wavelength zero-points 
were corrected using the OI line at 6300.304 \AA, in some cases by up to 
$\sim5$ \AA.  No skylines are included in the blue spectra, making it 
difficult to test whether shifts in the zero-point of the wavelength scale 
are present between the individual exposures, and the spectra extracted from 
individual exposures were too faint to test for such differences by 
cross-correlation. However, we extracted spectra of the underlying galaxy 
light and cross-correlated these to test for any shifts in the wavelength 
scale, and found no systematic differences between individual exposures 
larger than 1\AA.

  Data for the observed cluster candidates in NGC~1023 and NGC~3384 are
listed in Tables~\ref{tab:spec1023} and \ref{tab:spec3384}. Radial velocities
were measured by cross-correlating the object spectra with spectra of
three stars of spectral types G0--K3 \citep[numbers 17495, 33815 and 36094 in 
the recent compilation by][]{bar00}. The
cross-correlation was done with the FXCOR task in the RV task in IRAF.
Whenever possible, we measured radial velocities on both the blue and
red spectra. 
The tables list all candidate clusters included in the LRIS slitmasks, 
including a few for which no significant cross-correlation peak was
found. Some of these have very poor S/N, and for others the delicate 
extraction of the faint spectra was compromised because of internal 
reflections (``ghosts'') in the blue arm of the spectrograph or uncertain 
background subtraction because the object was located near the end of a 
slitlet.  It can, of course, not be ruled out that some of these objects 
are indeed background galaxies and that this is the reason why no 
cross-correlation
signal was found anywhere near the expected radial velocity. However,
it is clear that the vast majority of the objects have radial velocities 
indicating association with NGC~1023 or NGC~3384. 

  The unweighted means of the cluster velocities for NGC 1023 and NGC 3384,
measured on the blue spectra, are $692\pm56$ km/s and $1149\pm43$ km/s
respectively, in both cases larger than the RC3 values for NGC~1023 (601
km/s) and NGC~3384 (704 km/s). However, for the clusters with radial
velocity measurements on both blue and red spectra, the mean differences
are $\langle V_b - V_r \rangle = 133\pm31$ km/s for NGC~1023 and $\langle
V_b - V_r \rangle = 381\pm54$ km/s for NGC~3384.  If all the radial
velocities measured on the blue spectra are corrected for these differences
then the mean values become $559\pm64$ km/s and $768\pm79$ km/s,
respectively, in much better agreement with the cataloged values for the
parent galaxies.

\section{The characteristics of the faint extended clusters}

\subsection{Uniqueness}

Are these objects really something new?  After all, it is well known that
the Milky Way hosts several diffuse faint clusters and there is at least one
(NGC 2257) in the LMC \citep{ef85}.  In Fig.~\ref{fig:brodie.fig2} we show the
absolute magnitudes of Milky Way globular clusters as a function of their
metallicities, where the data are from \citet{har96}. 
We used [Fe/H] rather than color here because metallicity is
available for more clusters than $V-I$ or any other color. As in the 
lower panels of Fig.~1 , the sizes of the
points in this plot correspond to the half-light radii of the clusters. The
box in this figure corresponds to the selection box in the lower left panel
of Fig.~\ref{fig:brodie.fig1}. Clearly, there are no Milky Way objects
which correspond in metallicity (color), luminosity and size to the faint
extended clusters in NGC 1023. There are a few Milky Way objects
which have similar sizes and luminosities but their metallicities are
extremely different. This is true too of the extended cluster in the LMC.
The group of ``Palomar-Like'' Milky Way globular clusters are clearly
present as a small group of lower-luminosity blue objects. Such objects
might also be present in NGC 1023 but would be below the limit where we can
reliably measure sizes.

NGC 1023 has 29 objects with half-light radii, R$_h$, greater than 7 pc and
V$<$24 (i.e. M$_V$ brighter than --6.2, our limit for secure size
measurement).  According to \citet{har96}, the Milky Way has 19 globular 
clusters
larger than 7 pc but only 6 of these are brighter than M$_V$=--6.2. All
clusters with R$_h$ $>$7 pc in the Milky Way are located more than 4 kpc
from the plane of the Galactic disk and all but one are located at 20--80
kpc from the center of the Galaxy (the remaining one is at 7 kpc from the
center). The extended clusters in the Milky Way are clearly not associated
with its disk but are rather outer halo objects, presumably formed in the
low density environment in the outskirts of the protogalaxy \citep{mcl00}. 

We note in passing that Milky Way open clusters, contrary to common
misconception, are not particularly large, except for those which are
extremely young ($\le$10 Myr) and presumably unbound. Milky Way open
clusters are, in any case, significantly less massive than the faint
extended clusters in NGC 1023 \citep{jan88}.

\subsection{System Rotation}

From their mean velocities it is clear that the faint extended clusters are
associated with their respective galaxies and do not constitute a
background population.  Moreover, in Fig.~\ref{fig:brodie.fig3}, the radial
velocity plot for the faint extended clusters in NGC 1023, we see clear
evidence for rotation of the cluster system and this kinematic signature
corresponds quite closely to the rotation of the galaxy itself, as measured
along the major axis by \citet{sp97}.  The cluster radial velocities were
measured from the blue spectra and have been corrected by $-133$ km/s as
determined in Section~\ref{sec:data}.  The bulge effective radius for NGC
1023 is $<$2 kpc \citep{mh01} so the clusters are $\sim$2 bulge effective
radii from the galaxy center (1\arcmin=2.9 kpc at the distance of NGC
1023). This strongly suggests that these objects are associated with the
galaxy's disk rather than its bulge.

In Fig.~\ref{fig:brodie.fig4} we show, for comparison, the radial velocities 
of the compact globular clusters in NGC 1023, taken from \citet{lb02}. 
Although the sample is small, there is no hint of rotation in the compact 
cluster system.

Although there are too few clusters to provide any dynamical constraints on
the NGC~3384 extended cluster system, in Fig.~\ref{fig:brodie.fig5} we
show, for completeness, the radial velocities of the extended clusters and
the NGC~3384 rotation curve from \citet{fis97}.

\subsection{Ages and Metallicities}

The spectra of these extended objects are very similar to those of
moderately metal-rich old globular clusters.  Metallicities were determined
from a weighted combination of line strength indices according to the
prescription of \citet{bh90}. Because of the low S/N of the individual
spectra, a ``master'' spectrum (blue: Fig.~\ref{fig:brodie.fig6}, red:
Fig.~\ref{fig:brodie.fig7}) was produced by co-adding all spectra with
known radial velocities, shifted to 0 velocity. For 14 clusters in NGC
1023, each with an exposure time of 9.5 hours, the ``master'' spectrum
represents a total of 133 hours of 10-meter telescope time. This co-added
spectrum has a S/N of 17 per pixel (i.e.~$\sim$34 per resolution element)
and yields a mean $\feh = -0.58\pm0.24$.  The average reddening-corrected
$V-I$ color ($<V-I>_o=1.22$) corresponds to $\feh = -0.51$, using the
color-metallicity conversion relation of \citet{kis98}, in excellent
agreement with the spectroscopic value.  Index values and metallicities
derived from the ``master'' spectrum are given in table~\ref{tab:ind1023}.

Values of $\feh$ and H$\beta$ measured on the NGC 1023 ``master'' spectrum
were compared to \citet{mt00} stellar evolutionary models
(Fig~\ref{fig:brodie.fig8}).  This yields only loose constraints on cluster
ages which are most probably $\sim$13 Gyr old, but ages as young as
$\sim$7--8 Gyr are not ruled out. The lower limit on H$\beta$ is well below
any model predictions so our data do not put any constraints on the upper age 
limit of the clusters.

The fact that these objects are old suggests that they are quite stable
against disruption. It might indicate that they are on roughly circular
orbits as this would be likely to minimize disruptive effects (disk/bulge
shocking, etc).

A similar summation exercise for NGC~3384, excluding N3384-FF-7, which is
over a magnitude brighter than the other extended clusters and which may be
a ``normal'' (but large) globular cluster, produces a co-added spectrum
with S/N = 9 per pixel. For these 5 clusters, each with 5 hours of
integration, the equivalent exposure time of the co-added spectrum is 25
hours. It yields a mean $\feh$ estimate of $-1.13\pm0.45$. The average
reddening-corrected $V-I$ color ($<V-I>_o=1.27$) corresponds to $\feh =
-0.35$, using the \citet{kis98} relation. If N3384-FF-7 is included, the
S/N increases to 11 and $\feh = -0.64\pm0.34$. The mean color is
unchanged. Because of the large errors on $\feh$ and H$\beta$, no useful
constraints on age can be obtained even from the summed
spectrum. Table~\ref{tab:ind3384} gives the relevant indices from the NGC
3384 ``master'' spectrum. TiO$_{12.5}$ is not included because a red summed
spectrum was not formed for this galaxy due to poor signal-to-noise.

\subsection{$\alpha$-element to iron ratios}

Relations between Mgb and [Fe/H] (Fig~\ref{fig:brodie.fig9}) and
TiO$_{12.5}$ and [Fe/H] (Fig~\ref{fig:brodie.fig10}) for various levels of
$\alpha$-element enhancement were produced from the SSP models of
\citet{mbs00}, assuming a Salpeter IMF. For comparison with our cluster
measurements, the Mg2 values provided in \citet{mbs00} were converted to
Mgb using the relation $Mgb = 14.03 \times Mg2 + 0.361$, derived from the
Maraston models as in \citet{lar02}. This conversion is accurate to
$\le$0.1\AA\ for ages between 10 and 15 Gyr. $\feh$ is used rather than
$<Fe>$ because the latter is an average of two Fe indices Fe5270 and
Fe5335, the redder of which is close to the edge of our spectral range and
is poorly determined in our data. We note, however, that since Mg is a
component in the $\feh$ determination, if the objects are
$\alpha$-enhanced, the resultant $\feh$ value will be too high. Data points
from the NGC 1023 ``master'' spectrum indicate that the [$\alpha$/Fe] ratio
for the faint extended objects is between +0.3 and +0.6. Any adjustment for
the inflated $\feh$ values will move points to the left in the plots and
will tend to further increase the $\alpha$-enhancement. Also shown in
Fig~\ref{fig:brodie.fig9} is the relation from \citet{mar01} for a 12 Gyr
SSP model. Since
many of the metal-poor stars used in deriving the Maraston relation may be 
$\alpha$-enhanced \citep{mt00},
its agreement
with the \citet{mbs00} $\alpha$-enhanced curves is not surprising.
[$\alpha$/Fe] ratios of $\sim$+0.3 are typical for old stellar populations
in the bulge and halo of the Milky Way \citep{car96} and the globular
clusters and starlight of elliptical galaxies \citep{for01, mbs00} and are
interpreted as indicating rapid, early chemical element enrichment
dominated by Type II supernovae.

\section{Discussion}

Observational evidence \citep{lar01, bur01, kw01} and current ideas about
the origins of globular cluster sub-populations \citep{az92, fbg97, vdb01}
suggest that the origins of {\it compact} red (metal-rich) and blue
(metal-poor) globular clusters may have been quite different, occurring at
different epochs or under different physical mechanisms, or both.  


A striking difference between the blue and red subpopulations of compact
globular clusters, which is especially interesting
in regard to these faint extended red clusters, is that within the
``normal'' compact globular cluster population, the red globulars are
always $\sim$25\% {\it smaller} than their blue counterparts
\citep{lar01, kw98, puz99}. These considerations underscore the fundamentally different
nature of the population of red extended objects in these lenticular
galaxies.


Perhaps clues about the origins of these objects can be found by asking
what NGC~1023 and NGC~3384 have in common that differentiates them from NGC
3115.  NGC~1023 is the dominant member of a well-defined group of 15
galaxies and NGC~3384 is a member of the Leo I group \citep{gar93}. By
contrast, NGC 3115 is isolated except for a dwarf companion NGC 3115DW at a
distance of 5.5 arcmin ($\sim$17.5 kpc). Both NGC~1023 and NGC~3384 are
classified as SB0s while NGC 3115 is a highly bulge-dominated galaxy,
transitional in type between E7 and S01. In Paper I we speculated that the
faint extended cluster population in NGC~1023 might have originated in
encounters with dwarf companion galaxies.  According to \citet{hp94},
massive cluster formation occurs within supergiant molecular clouds that
are pressure-confined by the parent galaxy's interstellar
medium. Therefore, more extended clusters would naturally form in
lower-density environments like the outer regions of giant galaxies and the
weak potential wells of dwarf galaxies. Similarly, \citet{mcl00} has noted
a relation between cluster binding energy and galactocentric radius for
Milky Way globular clusters which implies extended cluster formation in low
ambient density environments. These considerations might lend support to
the idea that the faint extended clusters formed in dwarfs before joining
NGC 1023.  Young star clusters have formed recently in the nearby dwarf,
NGC~1023A \citep{lb02} but with no HST observations of this companion
galaxy, we cannot say whether these new clusters are extended.  It would be
natural to suppose that more dwarf companion galaxies were present in the
past in the vicinity of NGC~1023 and NGC~3384, given their presence in
significant groups.

Alternatively, the fact that NGC~3115 has such a modest disk may be the
relevant feature. It is difficult to preferentially destroy compact objects
while preserving extended ones so, if the faint extended clusters formed in
the disk, the ambient conditions (gas densities?) there may have been
conducive to forming only extended objects. Lenticulars with well-developed
disks may provide an environment for the formation of star clusters with 
unique characteristics.

Another factor may be the sizes of the black holes at the centers of these
galaxies.  NGC 3115 has a much more massive (9.1$\times10^8 M_\odot$) black
hole in its center than do NGC 1023 (3.9$\times10^7 M_\odot$) and NGC 3384
(1.8$\times10^7 M_\odot$), where the black hole masses are from
\citet{ho02}. In other words old clusters in NGC 3115 may have been
subjected to quasar-like environmental conditions during their youth,
whereas clusters in NGC 1023 and NGC 3384 might have spent their youths in
less stressful environments \citep{vdb02}.

\section{Conclusions}

We have demonstrated conclusively that the faint extended objects
discovered in HST images of NGC~1023 and NGC~3384 are star clusters
associated with their respective galaxies. In the case of NGC~1023 we were
further able to establish that these objects are old, moderately metal-rich
and, having a system rotation which is very similar to the rotation curve
of the host galaxy, are associated with the lenticular disk.  

As a population, these star clusters have no known analog in the Milky Way
or indeed elsewhere in the Local Group. Comparisons with the Milky Way
``Palomar-like'' extended clusters confirms that the Milky Way objects are
similar only in size. They differ in number, kinematics, location with
respect to the disk and the Galaxy center, metallicity and luminosity. In
short, the Milky Way objects are a small halo population of metal-poor,
low-luminosity clusters while the NGC 1023 cluster are numerous,
metal-rich, disk objects which rotate as an ensemble.

 These new members of the cluster family have been discovered in two
lenticular galaxies and the presence of similar objects has been definitely
ruled out in a third lenticular as well as in a nearby elliptical
galaxy. Existing HST data for other galaxies are inadequate to establish
either the presence or the absence of such clusters so, without further
observations, we cannot ascertain how common this phenomenon might be and
whether or not these newly-discovered objects are found exclusively in
lenticular galaxies. If so, they are not found in {\it all}
lenticulars. Late-type S0s may offer a more suitable environment for the
formation of extended clusters than early-type lenticulars. The size of the
host galaxy's central black hole may be a governing factor in allowing an
appropriate environment for the survival of extended disk clusters. If
galaxy-galaxy interactions play a role in forming these clusters, group (or
galaxy cluster) membership may be a relevant criterion for selecting host
galaxy candidates.

\acknowledgments

We thank Sidney van den Bergh, Duncan Forbes and an anonymous referee for
helpful comments and suggestions. This work was supported by National
Science Foundation grant number AST-9900732.

\onecolumn

\begin{deluxetable}{lllccccr}
\tablecaption{ \label{tab:spec1023}
  Spectroscopic data for cluster candidates in NGC~1023}
\tablecomments{FF = ``Faint Fuzzies'' our informal name for the faint extended clusters, RV = heliocentric radial velocity, $V$ = visual magnitude.
 The S/N column gives the average signal-to-noise per pixel on the blue
 spectra in the region 4000\AA -- 5000 \AA . The photometry has not been 
 corrected for reddening.
}
\tablehead{Object   & RA(2000)   & DEC(2000)
	           & $V$      & \vi & \multicolumn{2}{c}{RV (km/s)}  & S/N \\
	            &            &
                   &                &               &  blue    &    red  & \\
		   }
\startdata
NGC 1023     & 2:40:24.1  & 39:03:46   &        -       &       -       &  \multicolumn{2}{c}{601} &  -  \\
N1023-FF-1 & 2:40:34.73 & 39:02:58.6 & $23.60\pm0.03$ & $1.19\pm0.04$ & $1011\pm48$ &    -       & 3.6 \\ 
N1023-FF-2 & 2:40:33.79 & 39:04:31.2 & $23.29\pm0.02$ & $1.39\pm0.03$ & $994\pm9$   &    -       & 5.0 \\ 
N1023-FF-3  & 2:40:32.79 & 39:04:10.8 & $23.62\pm0.03$ & $1.39\pm0.03$ &    -        &    -       & 2.1 \\ 
N1023-FF-4  & 2:40:31.94 & 39:03:42.9 & $23.82\pm0.03$ & $1.20\pm0.04$ & $736\pm40$  &    -       & 2.2 \\ 
N1023-FF-5  & 2:40:29.96 & 39:03:36.3 & $22.56\pm0.01$ & $1.27\pm0.02$ & $1015\pm35$ & $888\pm6$  & 8.6 \\ 
N1023-FF-6  & 2:40:29.34 & 39:03:17.2 & $23.65\pm0.03$ & $1.33\pm0.04$ & $811\pm78$  & $663\pm21$ & 3.9 \\ 
N1023-FF-7  & 2:40:28.78 & 39:03:17.9 & $23.46\pm0.02$ & $1.36\pm0.03$ & $666\pm73$  & $565\pm4$  & 4.5 \\ 
N1023-FF-8 & 2:40:28.20 & 39:03:47.1 & $23.68\pm0.03$ & $1.39\pm0.03$ &    -        &    -       & 2.6 \\ 
N1023-FF-9 & 2:40:27.50 & 39:04:15.7 & $23.40\pm0.02$ & $1.28\pm0.03$ & $648\pm98$  &    -       & 3.5 \\ 
N1023-FF-10 & 2:40:26.40 & 39:04:19.6 & $23.31\pm0.02$ & $1.31\pm0.03$ & $614\pm78$  & $629\pm66$ & 6.6 \\ 
N1023-FF-11 & 2:40:23.53 & 39:04:24.4 & $23.71\pm0.03$ & $1.36\pm0.04$ &    -        &    -       & -  \\ 
N1023-FF-12  & 2:40:21.55 & 39:04:16.6 & $22.99\pm0.02$ & $1.30\pm0.02$ & $514\pm8$   & $394\pm17$ & 6.9 \\ 
N1023-FF-13 & 2:40:20.02 & 39:02:55.1 & $23.79\pm0.03$ & $1.18\pm0.04$ & $571\pm30$  &    -       & 2.5 \\ 
N1023-FF-14  & 2:40:19.58 & 39:04:36.7 & $23.23\pm0.02$ & $1.34\pm0.03$ & $725\pm17$  & $489\pm45$ & 7.7 \\ 
N1023-FF-15  & 2:40:18.62 & 39:02:55.7 & $23.56\pm0.02$ & $1.40\pm0.03$ & $588\pm31$  & $322\pm6$  & 4.0 \\ 
N1023-FF-16  & 2:40:17.72 & 39:02:52.3 & $23.29\pm0.02$ & $1.01\pm0.03$ & $351\pm25$  &    -       & 6.5 \\ 
N1023-FF-17 & 2:40:16.92 & 39:04:04.2 & $23.16\pm0.02$ & $1.22\pm0.03$ &    -        &    -       & 6.5 \\ 
N1023-FF-18  & 2:40:16.35 & 39:03:29.0 & $22.68\pm0.01$ & $1.28\pm0.02$ & $438\pm38$  & $355\pm12$ & 7.0 \\ 
N1023-FF-19 & 2:40:15.66 & 39:03:48.1 & $23.49\pm0.02$ & $1.37\pm0.03$ &    -        &    -       &  -  \\ 
\enddata
\end{deluxetable}

\begin{deluxetable}{lllccccr}
\tablecaption{ \label{tab:spec3384}
  Spectroscopic data for cluster candidates in NGC~3384}
\tablecomments{FF = ``Faint Fuzzies'', our informal name for the faint extended clusters, RV = heliocentric radial velocity, $V$ = visual magnitude.
 The S/N column gives the average Signal-to-Noise per pixel in the region
 4000\AA -- 5000 \AA .
}
\tablehead{Object   & RA(2000)   & DEC(2000)
	           & $V$      & \vi & \multicolumn{2}{c}{RV (km/s)}  & S/N \\
	            &            &
                   &                &               &  blue    &    red  & \\
		   }
\startdata
NGC 3384     & 10:48:17.2  & 12:37:49   &      -         &       -       & \multicolumn{2}{c}{704}     &  -  \\
N3384-FF-1  & 10:48:17.50 & 12:38:29.7 & $23.28\pm0.09$ & $1.15\pm0.12$ &   -          &     -      & 4.7 \\  
N3384-FF-2   & 10:48:16.39 & 12:37:01.4 & $23.66\pm0.12$ & $1.35\pm0.14$ &    -         &     -      & 1.4 \\ 
N3384-FF-3  & 10:48:15.23 & 12:36:45.2 & $22.75\pm0.06$ & $1.30\pm0.06$ & $1100\pm 49$ & $609\pm48$ & 3.9 \\  
N3384-FF-4   & 10:48:15.00 & 12:39:10.9 & $23.60\pm0.12$ & $1.24\pm0.17$ & $1299\pm 21$ &     -      &  9.9 \\  
N3384-FF-5   & 10:48:14.80 & 12:39:02.4 & $23.68\pm0.13$ & $1.29\pm0.15$ &   -          &     -      & 2.8 \\ 
N3384-FF-6   & 10:48:14.16 & 12:37:34.2 & $22.87\pm0.06$ & $1.36\pm0.07$ & $1011\pm 15$ &     -      & 4.8 \\ 
N3384-FF-7   & 10:48:13.99 & 12:37:14.3 & $21.51\pm0.02$ & $1.26\pm0.03$ & $1151\pm  3$ & $905\pm23$ & 15.8 \\ 
N3384-FF-8   & 10:48:13.73 & 12:38:29.3 & $23.81\pm0.14$ & $1.25\pm0.17$ &   -          &     -      & 1.2 \\  
N3384-FF-9  & 10:48:12.72 & 12:36:29.3 & $23.29\pm0.09$ & $1.24\pm0.11$ & $1094\pm 42$ & $654\pm27$ & 6.5 \\  
N3384-FF-10  & 10:48:12.33 & 12:36:45.3 & $23.32\pm0.09$ & $1.19\pm0.11$ & $1242\pm 24$ & $895\pm30$ & 1.9 \\  
N3384-FF-11  & 10:48:10.46 & 12:38:29.7 & $23.60\pm0.12$ & $1.34\pm0.14$ &    -         &     -      &  2.8 \\ 
\enddata
\end{deluxetable}

\begin{deluxetable}{lccc}
\tablecolumns{4}
\tablewidth{0pc}
\tablecaption{ \label{tab:ind1023}
  Indices measured from the NGC 1023 summed spectrum}

\tablecomments{References: (1) \citet{bh90}, (2) \citet{wor94}, (3)
\citet{mbs00}. Brodie \& Huchra indices are measured in magnitudes, the
rest are equivalent widths in \AA.  The $\feh$ values are metallicities
derived from each individual index strength. The final quoted metallicity
is a weighted mean of the individual index measurements as in
\citet{bh90}.}

\tablehead{Index & Mag./\AA & $\feh$ & Ref. \\}
\startdata
Delta  & 0.627$\pm$0.019 & -0.44$\pm$0.37 & 1 \\
Mg2    & 0.215$\pm$0.013 & -0.07$\pm$0.36 & 1 \\
MgH    & 0.111$\pm$0.010 & ~0.45$\pm$0.52 & 1 \\
Gband  & 0.126$\pm$0.028 & -1.02$\pm$0.44 & 1 \\
CNB    & 0.101$\pm$0.029 & -1.27$\pm$0.38 & 1 \\
Fe5270 & 0.073$\pm$0.010 & -0.60$\pm$0.64 & 1 \\
CNR    & 0.070$\pm$0.020 & -0.68$\pm$0.47 & 1 \\
HK     & 0.301$\pm$0.029 & -0.83$\pm$0.48 & 1 \\
H$\beta$ & 1.868$\pm$0.477 &  & 2 \\
Mgb & 3.277$\pm$0.322 &  & 2 \\
TiO$_{12.5}$ & 22.340$\pm$4.637 & & 3 \\
\enddata
\end{deluxetable}

\begin{deluxetable}{lccc}
\tablecolumns{4}
\tablewidth{0pc}
\tablecaption{ \label{tab:ind3384}
  Indices measured from the NGC 3384 summed spectrum}

\tablecomments{References: (1) \citet{bh90}, (2) \citet{wor94}. Brodie \&
Huchra indices are measured in magnitudes, Worthey indices are equivalent
widths in \AA. The summed spectrum does not include N3384-FF-7. TiO$_{12.5}$
is not given for NGC 3384 because the signal-to-noise did not warrant
producing a red summed spectrum. The $\feh$ values are metallicities
derived from each individual index strength. The final quoted metallicity
is a weighted mean of the individual index measurements as in
\citet{bh90}.}

\tablehead{Index & Mag./\AA & $\feh$ & Ref. \\}
\startdata
Delta  & ~0.640$\pm$0.026 & -0.39$\pm$0.37 & 1 \\
Mg2    & ~0.137$\pm$0.025 & -0.82$\pm$0.42 & 1 \\
MgH    & ~0.085$\pm$0.021 & -0.04$\pm$0.64 & 1 \\
Gband  & -0.017$\pm$0.048 & -2.65$\pm$0.63 & 1 \\
CNB    & ~0.312$\pm$0.078 & ~0.11$\pm$0.61 & 1 \\
Fe5270 & ~0.040$\pm$0.027 & -1.27$\pm$0.82 & 1 \\
CNR    & -0.064$\pm$0.041 & -1.66$\pm$0.54 & 1 \\
HK     & ~0.181$\pm$0.063 & -1.79$\pm$0.66 & 1 \\
H$\beta$ & ~1.308$\pm$1.13 &  & 2 \\
Mgb & ~1.191$\pm$0.954 &  & 2 \\
\enddata
\end{deluxetable}

\clearpage
\epsfxsize=16cm \epsfbox{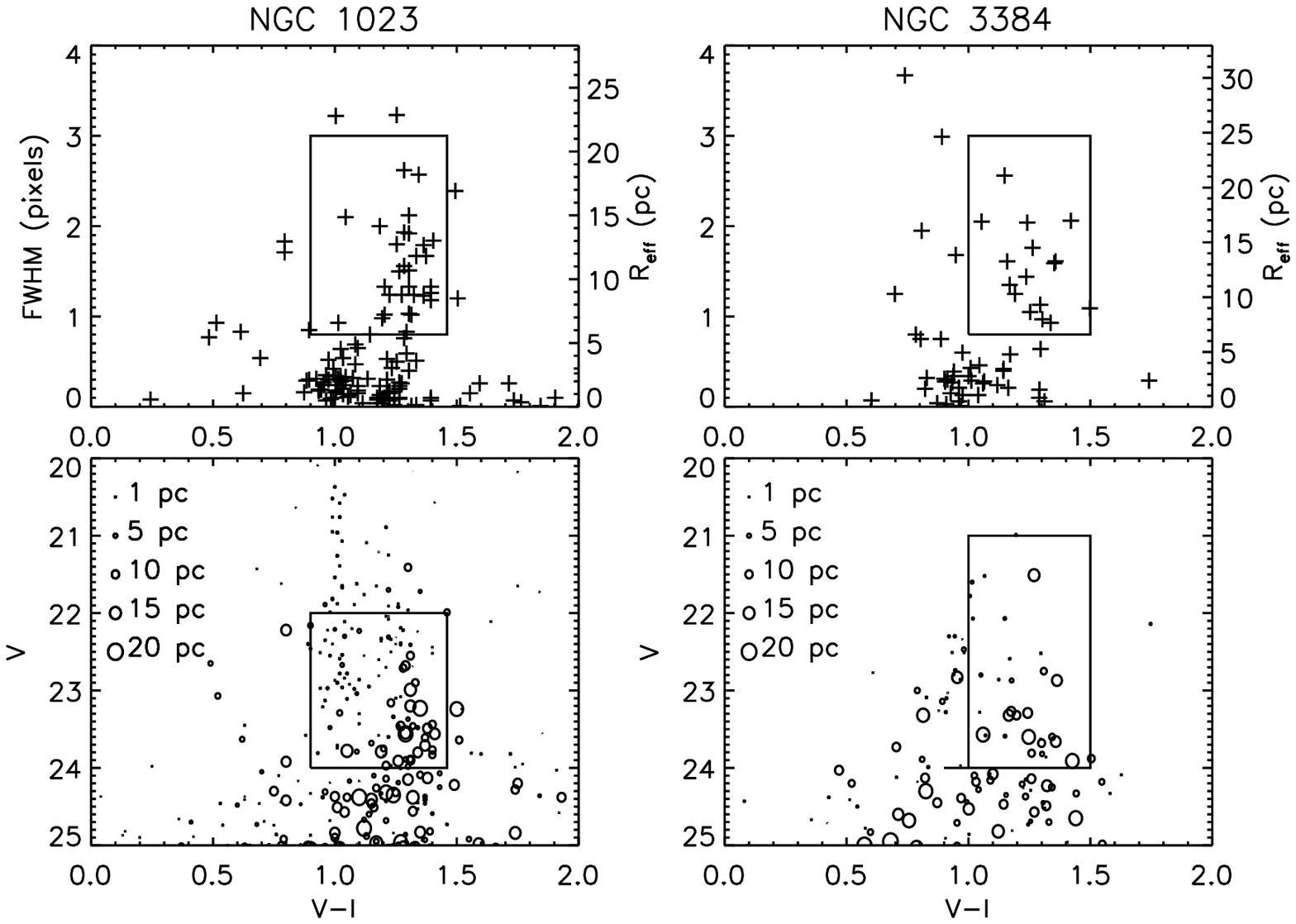} 
\figcaption{\label{fig:brodie.fig1}Selection
of cluster candidates for spectroscopy. The lower panels are
color-magnitude diagrams with symbol sizes proportional to the object sizes.
The upper panels show object size vs.\ \vi\ color for objects
brighter than $V=24$. The boxes indicate the selection criteria for objects
to be included in the Keck / LRIS slitmasks. }

\clearpage
\epsfxsize=14cm \epsfbox{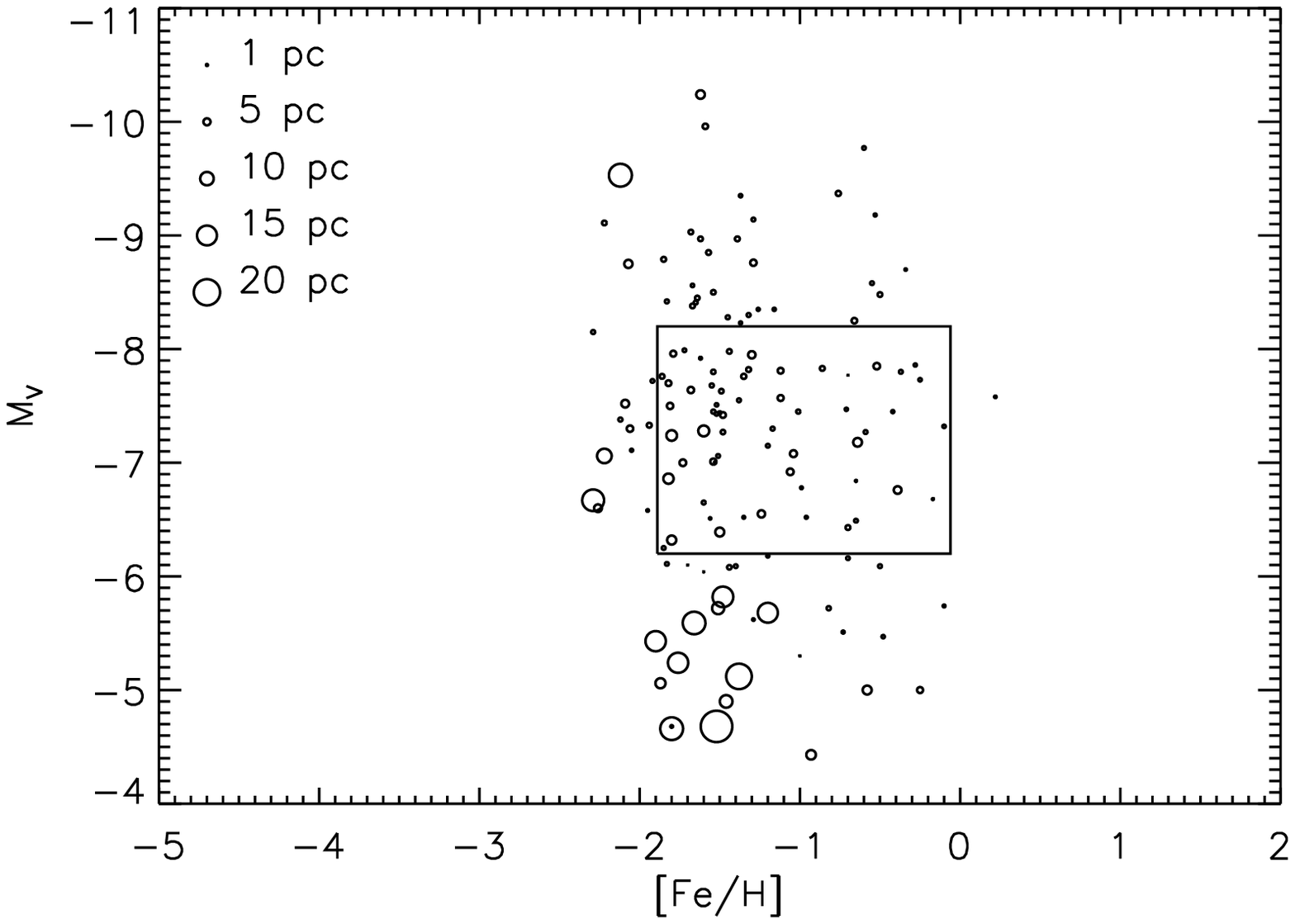}
\figcaption{\label{fig:brodie.fig2} Absolute V magnitude vs. [Fe/H] for
Milky Way globular clusters. The symbol sizes are proportional to the
object sizes. The box corresponds to the selection box in the lower left panel
of Figure 1.}

\clearpage
\epsfxsize=14cm
\epsfbox{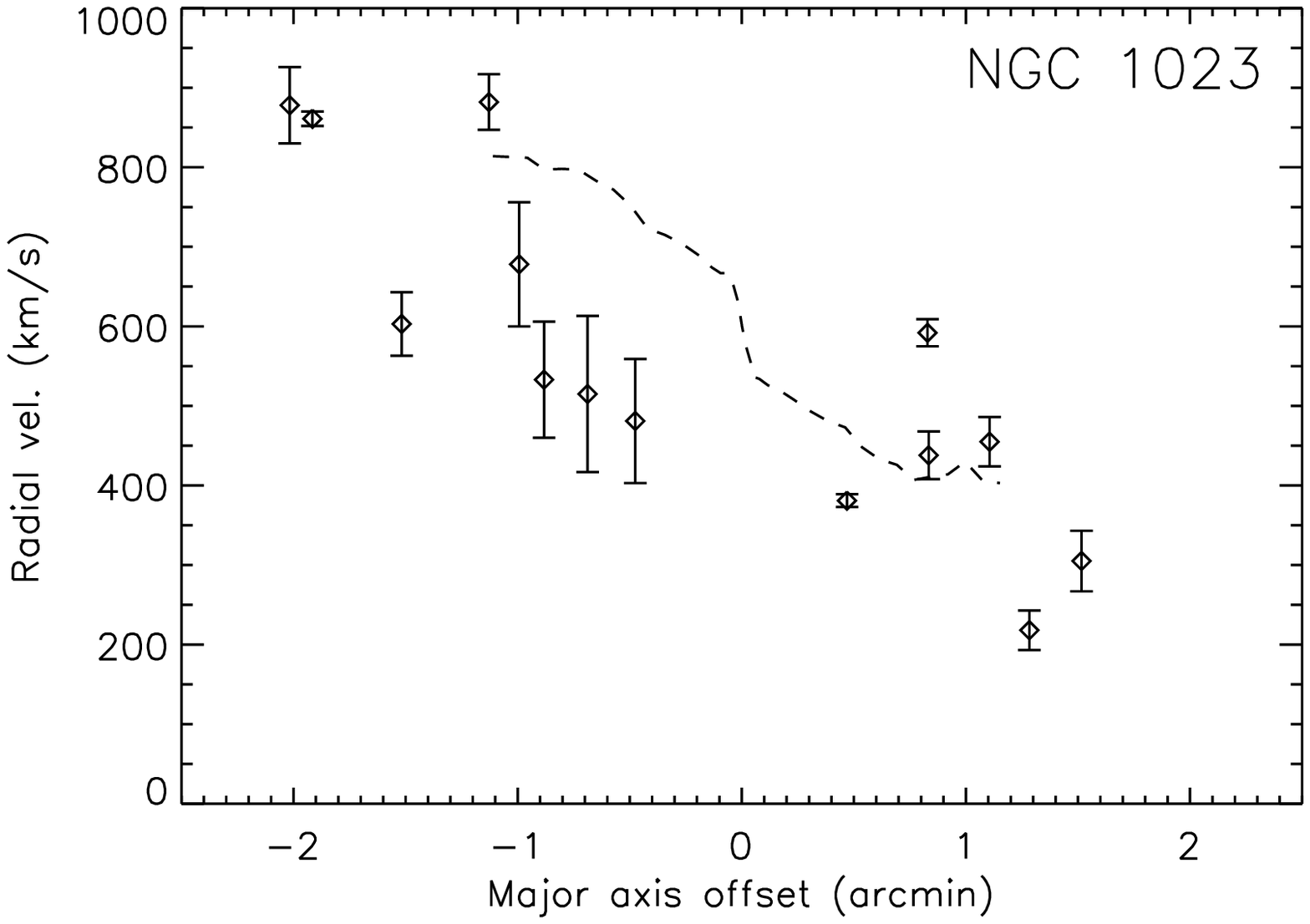}
\figcaption{\label{fig:brodie.fig3}Radial velocity vs.\ projected distance from
the galaxy center along the major axis for extended clusters in NGC~1023.
The radial velocities were measured from the blue spectra, applying a
correction of $-133$ km/s (see text).  The dashed line indicates the 
rotation curve for NGC~1023 itself from a longslit positioned along
the major axis, from \citet{sp97}.}

\clearpage
\epsfxsize=14cm
\epsfbox{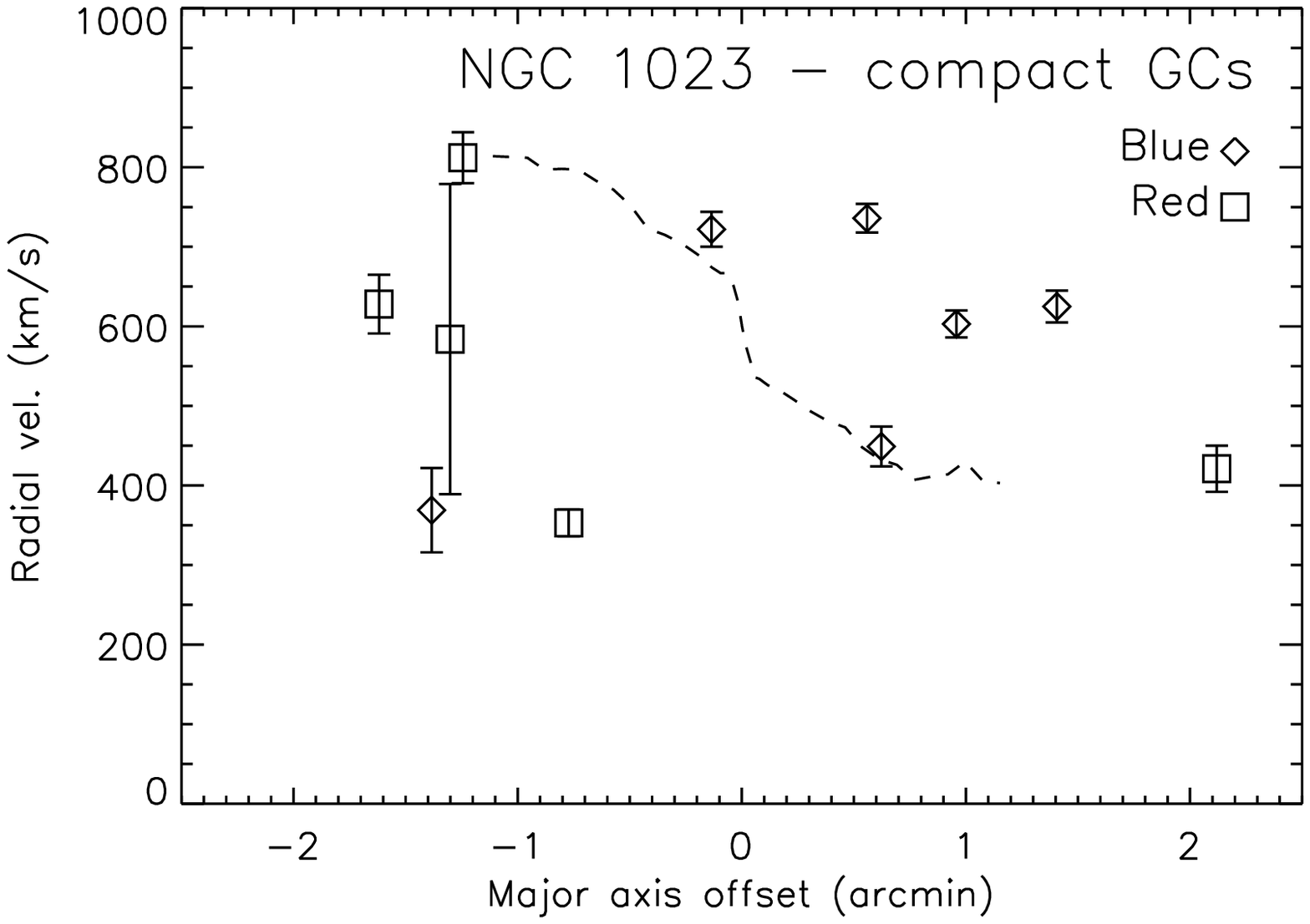}
\figcaption{\label{fig:brodie.fig4}Radial velocity vs.\ projected distance from
the galaxy center along the major axis for compact (globular) clusters 
in NGC~1023 from data in \citet{lb02}. The dashed line is the galaxy 
rotation curve as in Figure 3.}

\clearpage
\epsfxsize=14cm
\epsfbox{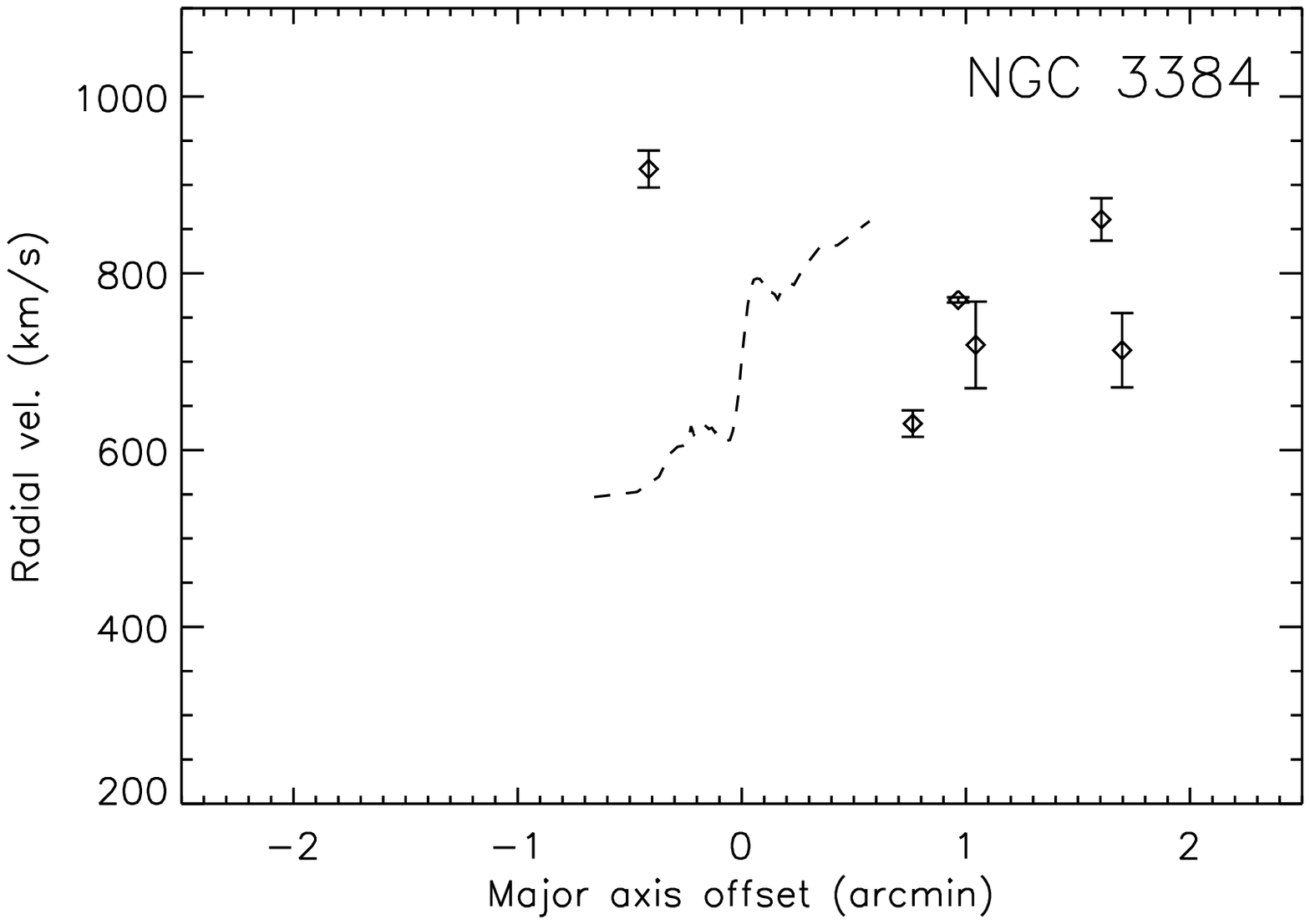}
\figcaption{\label{fig:brodie.fig5}Radial velocity vs.\ projected distance from
the galaxy center along the major axis for cluster candidates in NGC~3384.
Overplotted is the rotation curve for NGC~3384 from \citet{fis97}.}

\clearpage
\epsfxsize=14cm
\epsfbox{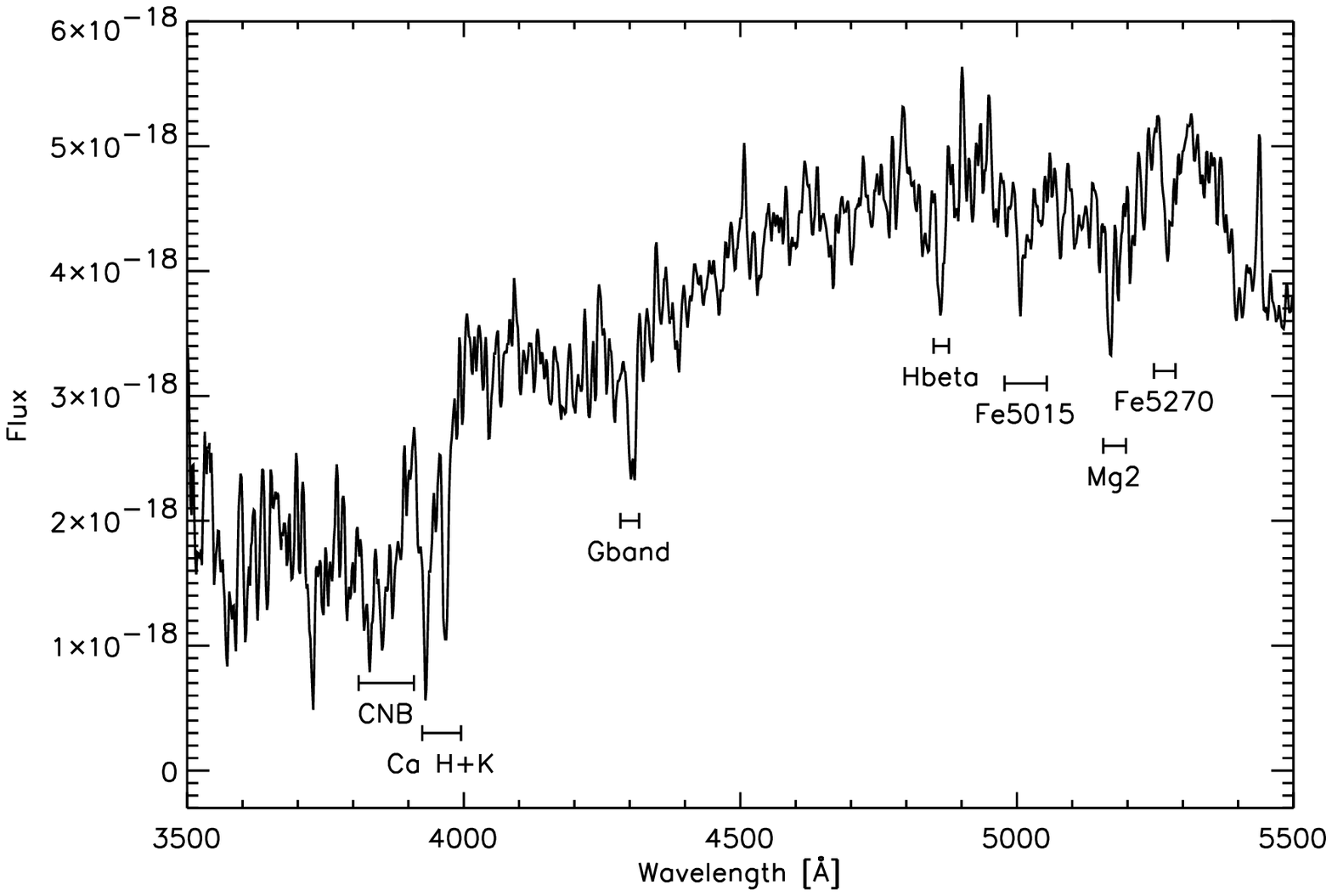}
\figcaption{\label{fig:brodie.fig6}Co-added blue spectrum of all clusters in
NGC~1023 with known radial velocities. The spectrum has been smoothed
with a 3 pixels boxcar filter. Various spectral features are indicated.}

\clearpage
\epsfxsize=14cm
\epsfbox{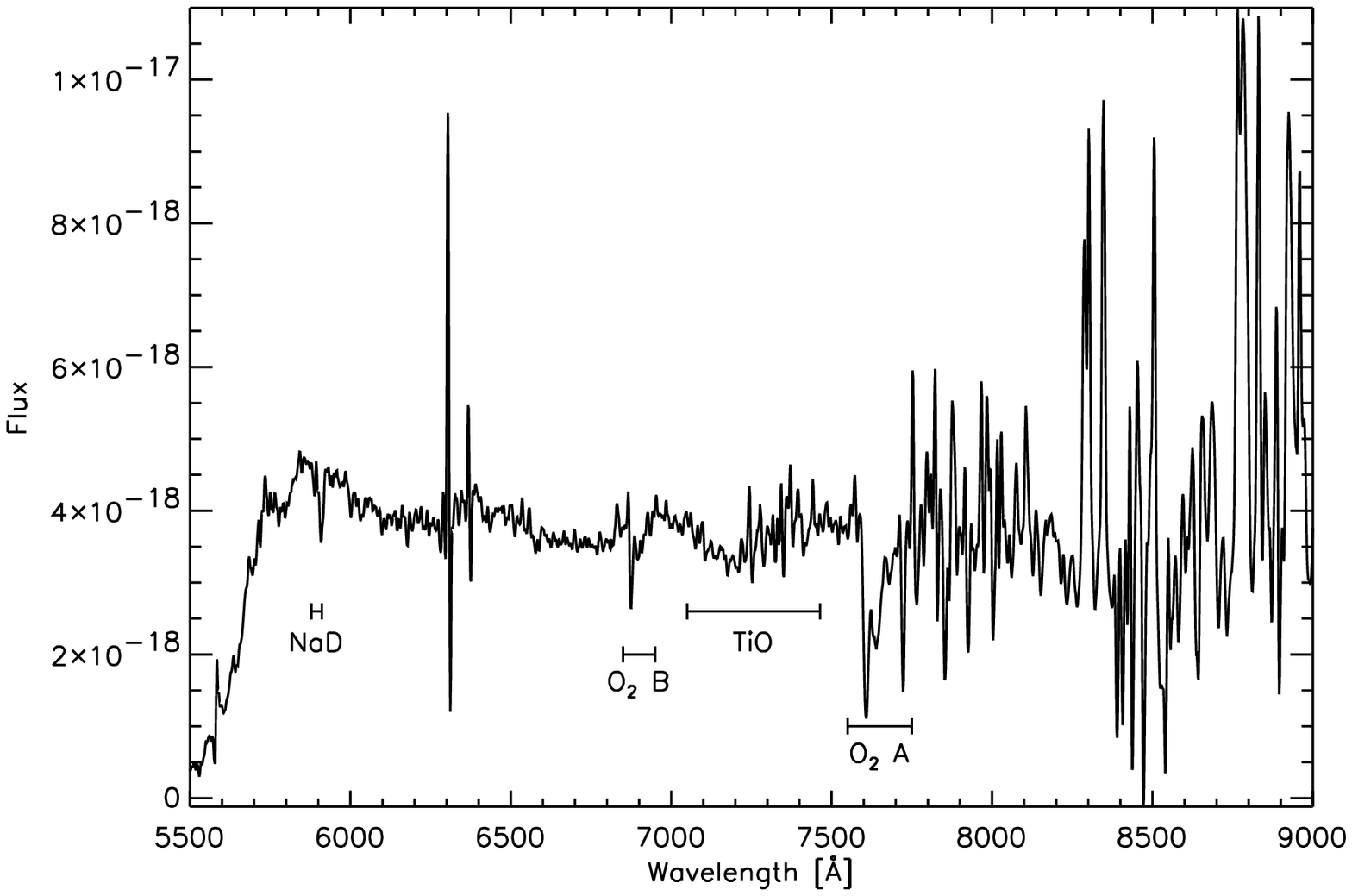}
\figcaption{\label{fig:brodie.fig7}Co-added red spectrum of all clusters in
NGC~1023 with known radial velocities. The spectrum has been smoothed
with a 3 pixels boxcar filter. Various spectral features are indicated,
including the terrestrial O$_2$ absorption bands.}

\clearpage
\epsfxsize=14cm 
\epsfbox{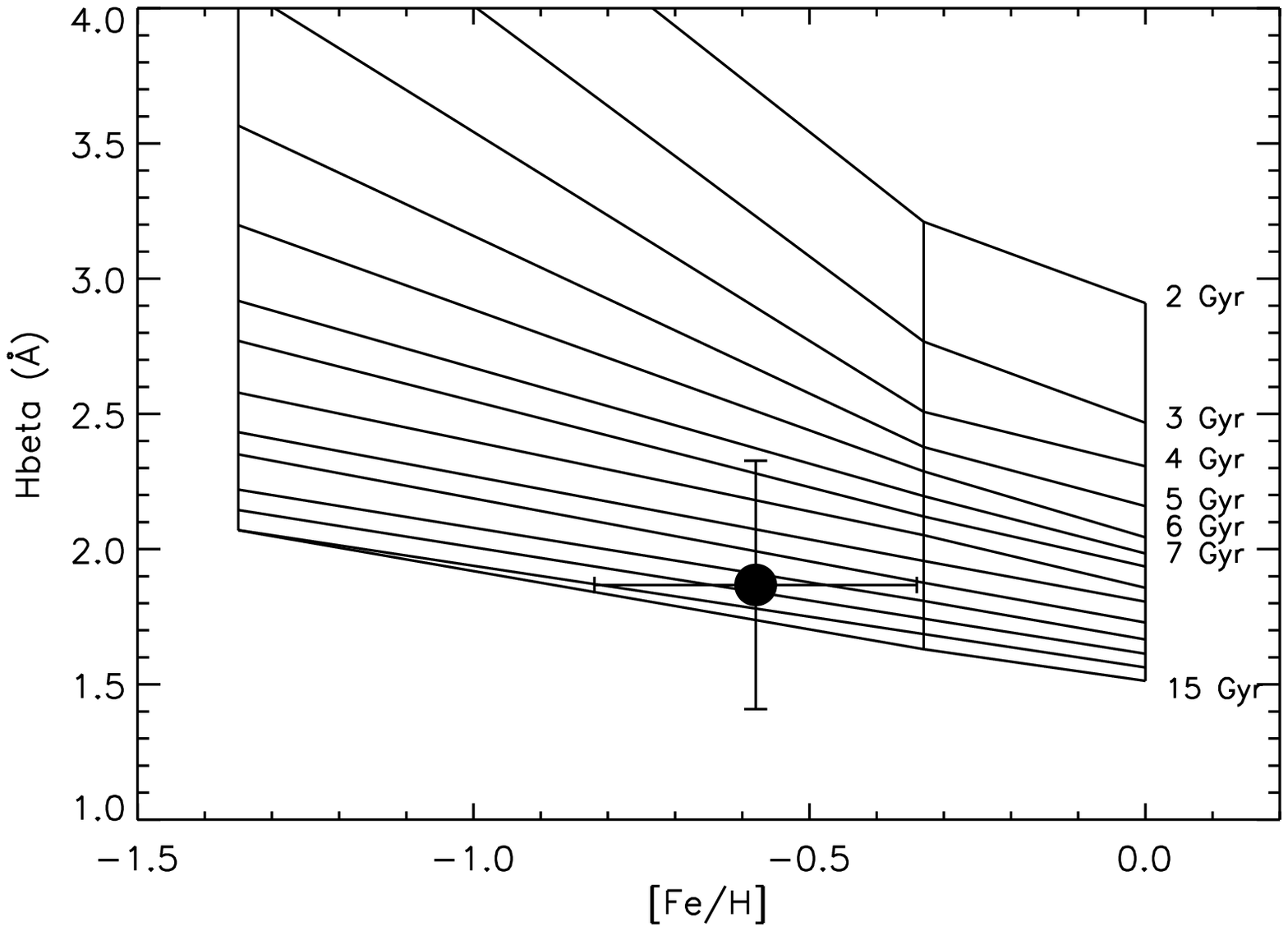} 
\figcaption{\label{fig:brodie.fig8}$\feh$
and H$\beta$ estimated from the co-added spectra of objects in NGC 1023
provide a loose constraint on the age of the clusters when compared to the
stellar evolutionary models of \citet{mt00}. The clusters are probably
$\sim$13 Gyr old but could be as young as $\sim$7--8 Gyr.}

\clearpage
\epsfxsize=14cm 
\epsfbox{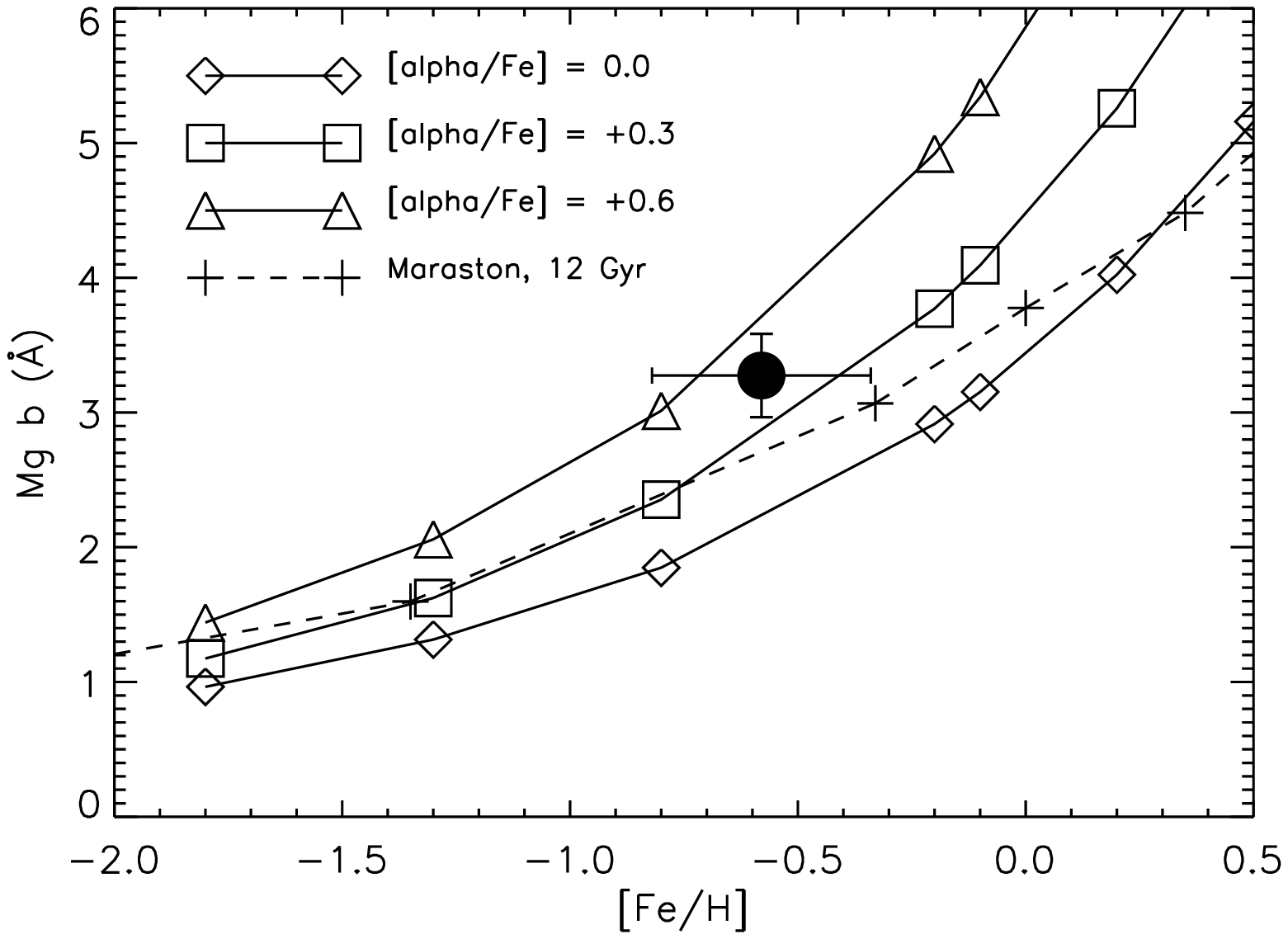} 
\figcaption{\label{fig:brodie.fig9}
$\feh$ and Mgb (solid dot) estimated from the co-added spectra of objects
in NGC 1023, compared to SSP model predictions from \citet{mbs00} for a
range of $\alpha$-element to iron enhancements, assuming
a Salpeter IMF.  Also shown is the \citet{mar01} SSP curve for a 12 Gyr
population.}

\clearpage
\epsfxsize=14cm 
\epsfbox{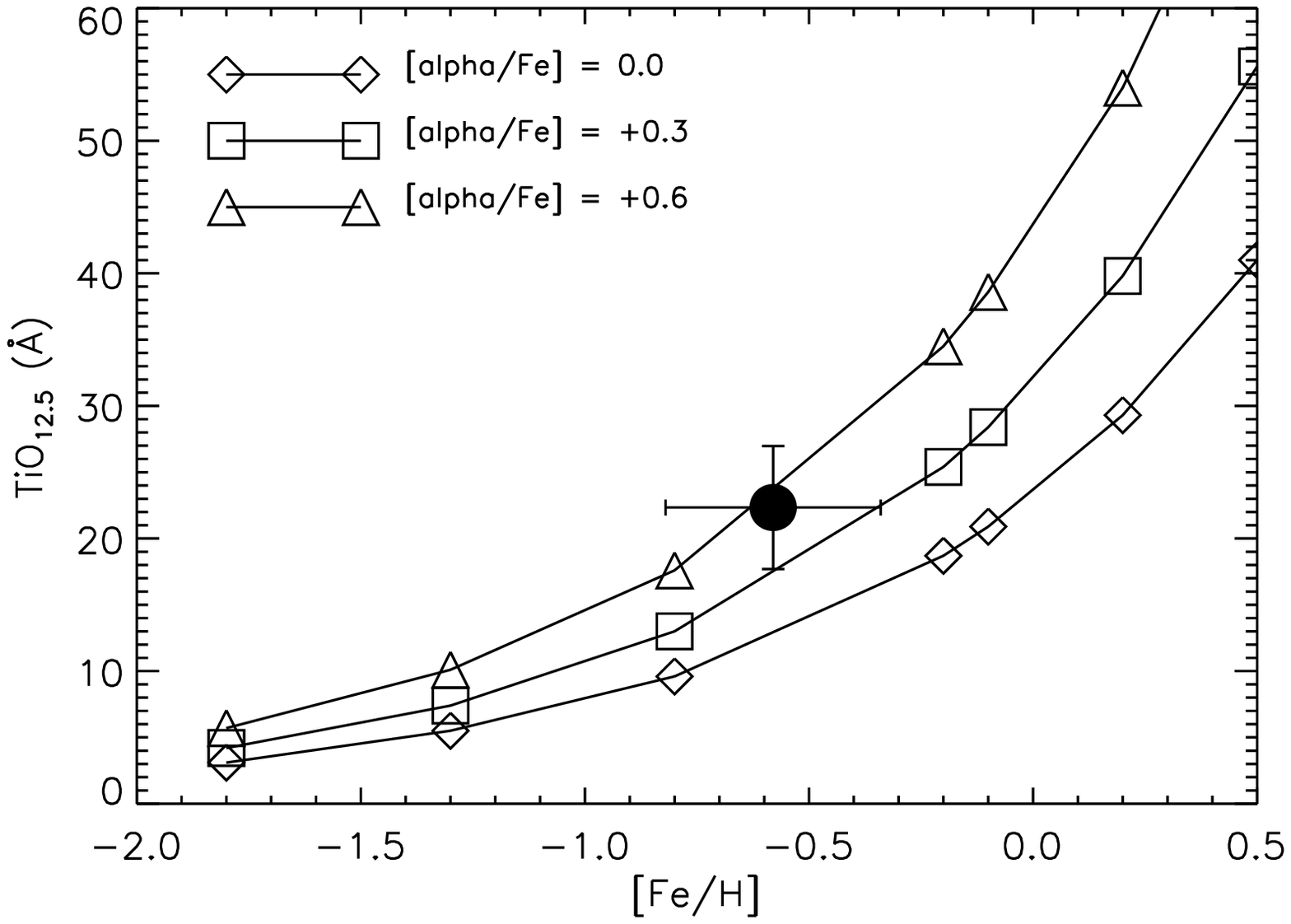} 
\figcaption{\label{fig:brodie.fig10}
$\feh$ and the TiO$_{12.5}$ index defined in \citet{mbs00} (solid dot)
estimated from the co-added spectra of objects in NGC 1023, compared to SSP
model predictions from \citet{mbs00} for a range of $\alpha$-element to
iron enhancements, assuming a Salpeter IMF.}
 
\end{document}